\documentclass{aa}

\usepackage{longtable,lscape}
\usepackage{graphicx}
\usepackage{txfonts}
\begin{document}

%

   \title{The Faber-Jackson relation for early-type galaxies: Dependence on the magnitude range}

   \titlerunning{The Faber-Jackson relation. Dependence on the magnitude range}

   \authorrunning{Nigoche-Netro et al.}

   \author{A. Nigoche-Netro  \inst{1}, J. A. L. Aguerri \inst{1,2}, P. Lagos \inst{1}, A. Ruelas-Mayorga \inst{3}, L. J. S\'anchez \inst{3}\and  A. Machado \inst{3} }


   \offprints{A. Nigoche-Netro}

   \institute{Instituto de Astrof\'isica de Canarias (IAC), V\'ia L\'actea s/n, 38200 La Laguna, Spain. \\ \email{anigoche;jalfonso;plagos@iac.es}
\and Departamento de Astrof\'isica, Universidad de La Laguna, C/ Molinos de Agua s/n, 38205, La Laguna, Spain.
\and Instituto de Astronom\'ia, Universidad Nacional Aut\'onoma de M\'exico, Apartado Postal 70-264,  Cd Universitaria 04510 \\
                M\'exico D.F., M\'exico.\\  \email{rarm;leonardo;machado@astroscu.unam.mx} }

  \abstract {} {Previous studies have found that the coefficients and intrinsic dispersions of both the Kormendy relation and the Fundamental Plane depend on the magnitude range within which the galaxies are contained. We study whether this type of behaviour is also present for the Faber-Jackson relation.} {We take a sample of early-type galaxies from the Sloan Digital Sky Survey (SDSS-DR7, $\sim$ 90 000 galaxies) spanning a range of approximately 7 $mag$ in both $g$ and $r$ filters and analyse the behaviour of the Faber-Jackson relation parameters as functions of the magnitude range. We calculate the parameters in two ways: i) We consider the faintest (brightest) galaxies in each sample and we progressively increase the width of the magnitude interval by inclusion of the brighter (fainter) galaxies (increasing-magnitude-intervals), and ii) we consider narrow-magnitude intervals of the same width ($\Delta M = 1.0$ $mag$) over the whole magnitude range available (narrow-magnitude-intervals).} {Our main results are that: i) in both increasing and narrow-magnitude-intervals the Faber-Jackson relation parameters change systematically, ii) non-parametric tests show that the fluctuations in the values of the slope of the Faber-Jackson relation are not products of chance variations.} {We conclude that the values of the Faber-Jackson relation parameters depend on the width of the magnitude range and the luminosity of galaxies within the magnitude range. This dependence is caused, to a great extent by the selection effects and because the geometrical shape of the distribution of galaxies on the $M - \log (\sigma_{0})$ plane depends on luminosity. We therefore emphasize that if the luminosity of galaxies or the width of the magnitude range or both are not taken into consideration when comparing the structural relations of galaxy samples for different wavelengths, environments, redshifts and luminosities, any differences found may be misinterpreted.}


   \maketitle

%

\section{Introduction}

\label{sec:intro}

The structural relations of early-type galaxies (ETGs) play an important role in the study of the formation and evolution of these galaxies. Among the most important structural relations for ETGs there are:
The Kormendy relation (KR; \cite{kor77})

\begin{equation}
\left\langle \mu\right\rangle_{e}=\alpha+\beta\log(r_{e}),
\end{equation}
the Faber-Jackson relation (FJR; \cite{fab76})

\begin{equation}
\log\sigma_{0}=A-BM_{},
\end{equation}
and the Fundamental Plane (FP; \cite{djo87}; \cite{dre87})

\begin{equation}
\log(r_{e})=a\log(\sigma_{0})+b\left\langle \mu\right\rangle _{e}+c.
\end{equation}
where $r_{e}$ represents the effective radius, $\left\langle \mu\right\rangle _{e}$ is the mean effective surface brightness inside  $r_{e}$, $M$ is the total absolute magnitude, $\sigma_0$ is the central velocity dispersion, and $\alpha$, $\beta$, $A$, $B$, $a$, $b$, and $c$ are scale factors.

The FP relation is a consequence of the dynamical
equilibrium condition (virial theorem) and the regular behaviour
of both the mass-luminosity ratio and the structure along the
entire range of ETGs luminosities (homology).  Based on these considerations, we should find that (a, b)=(2, -0.4). There are, however, discrepancies between these values and those obtained from observations. These discrepancies might indicate that the FP and other structural relations are not universal, i.e., that they may depend on wavelength, environment, redshift, and/or luminosity. There have been several studies of this universality using different samples of galaxies. We now discuss briefly the most important results about the universality of the structural relations.

Several studies have demonstrated that both of the FP coefficients $a$ and $b$ remain unchanged when the relation is examined at different wavelengths (\cite{ben92}; \cite{BENDERETAL98}; \cite{ber03b}; \cite{ber03c}; \cite{lab05}; \cite{lab08}). However, other studies have found that the structural relations do depend on wavelength (\cite{jor96}; \cite{hud97}; \cite{pah98}; \cite{sco98}; \cite{jun08}), and in particular, that the $a$ coefficient of the FP varies with wavelength (i.e., the coefficient is larger at longer wavelengths), while the $b$ coefficient remains stable, that is ($a$, $b$) $\sim$ (1.2-1.6, -0.34) for wavelengths of 0.55-8.0 $\mu m$. Some of the afore mentioned authors in this paragraph have interpreted this as an effect caused by colour gradients inside galaxies.

Several studies have also discovered that the structural relations and/or the structural parameters of galaxies are affected by the environment (\cite{ben92}; \cite{BENDERETAL98}; \cite{tru01}; \cite{tru02}; \cite{ber03b}; \cite{agu04}; \cite{gut04}; \cite{den05}; \cite{jor05}), although other studies have reached exactly the opposite conclusion (\cite{ros01}; \cite{tre01}; \cite{evs02}; \cite{gon03}; \cite{red04}; \cite{red05}; \cite{nig07}). This second set of results suggest that galaxies very quickly absorb the changes caused by gravitational interactions, not keeping memory of these changes in regard to the structural relations. Another possible explanation is that most of the perturbations affect only the external parts of galaxies which are not represented in the structural relations given that these are defined by the use of effective parameters only.

When the structural relations are constructed for galaxy samples at different redshifts, several studies report similarly (\cite{bar98}; \cite{zie99}; \cite{lab03}; \cite{bar06}) that only the zero point of these relations depends on the redshift and this dependence may be caused by the passive evolution of the stellar populations that form at high redshift ($z_{form} > 2$). Nevertheless, their results conflict with those of other authors. For example, Treu et al. (2005), J\o rgensen et al. (2006), and Fritz et al. (2009) find that the zero point of the FP depends on redshift but that the slope of the FP is steeper for higher redshift galaxies than for galaxies in the local Universe. They interpret this as a mass
dependence of the star formation history, that is, the low-mass galaxies ($10^{10.3}M_{\odot}$) have experienced star formation as recently as $z_{form} \sim 1.1$, while, galaxies with masses $ \sim 10^{10.8}M_{\odot}$ and masses $ > 10^{11.3}$ $M_{\odot}$ had their last major star formation episode at $z_{form} > 1.25$ and $z_{form} > 1.6$ (\cite{jor07}).

Finally, considering luminosity, Kormendy (1985) states that dwarf ellipticals do not follow the same plane as bright ellipticals. Other authors have reported similar results (\cite{ham87}; \cite{ben92}; \cite{cao93}; \cite{agu09}). Recent studies have demonstrated that some projections of the FP, such as the FJR depend on the mass, radial distance from the cluster centre (\cite{fri05}) and on the luminosity (\cite{des07}) in the sense that brighter galaxies have a steeper slope. These authors attribute these dependences to galaxies of different luminosities experiencing different formation histories and brighter galaxies being less affected by dissipation. Other studies indicate that it is inappropriate to draw conclusions about the physical properties of galaxies by comparing the coefficients of the structural relations for different magnitude ranges because the form of the distribution of galaxies in the space of the variables that define these relations plays a crucial role in determining their coefficients (\cite{nig07b}; \cite{nig08}; \cite{nig09}). These studies find that the detected changes in the coefficients values for the structural relations obtained from fits to samples of galaxies of different luminosity are not necessarily related to differences in the intrinsic properties of these galaxies.

The results presented in Nigoche-Netro (2007) and Nigoche-Netro et al. (2008; 2009) characterise the effects produced by the magnitude range restrictions for different galaxy samples and indicate that the differences between the coefficients calculated within a magnitude range may differ by as much as 60\% for the KR from the values calculated for another magnitude range,and up to 30\% for the FP. Using Monte Carlo simulations, these studies also show that the changes in the values of the coefficients for the KR and the FP relations may be attributed to a `geometrical effect', in other words, the geometrical form of the galaxy distribution in the variables space defining the KR and the FP relations changes systematically as brighter galaxies are considered. Therefore, the values of the coefficients also change, because the fitting of relations to galaxy distributions with different geometrical forms produces different results. An important conclusion of these studies is that any restriction imposed on any of the variables involved in the KR and FP relations will produce similar effects as those produced by the restrictions on the magnitude. In light of this, it is reasonable to assume that part of the discrepancies in results about the universality of the structural relations may have been caused by the large majority of studies having underestimated the effects produced by magnitude restrictions, as well as all restrictions affecting every variable involved in the determination of the parameters of the structural relations.

In this paper, we present a sample of ETGs from the SDSS-DR7 that contains approximately 90 000 galaxies and covers a relatively wide magnitude range ($\Delta M$ $\sim 7$ $mag$ in $g$ and $r$ filters). Using these data, we analyse the behaviour of the coefficients and the intrinsic dispersion of the FJR with respect to several characteristics of the magnitude range. Throughout this paper, we use $H_{0}$ = 70 $km s^{-1}Mpc^{-1}$, $\Omega_{m}$ = 0.3 and $\Omega_{\Lambda}$ = 0.7.

Our paper is organised as follows. In Sect. 2 we present the galaxy sample used to study the FJR. Section 3 presents the fitting method that we use in calculating the FJR coefficients, the results of these calculations and the analysis of the behaviour of the slope of the FJR as a function of the magnitude range. In Sect. 4 we present a discussion of the most important results of this paper. Finally, in Sect. 5 we present our conclusions.

\section{The sample of ETGs}
\label{sec:samps}

We extract a sample of ETGs from the Seventh Data Release of the SDSS (\cite{yor00}; \cite{aba09}) in $g$ and $r$ filters. This sample contains approximately 90 000 galaxies in each filter, distributed in a redshift interval $0.01 < z < 0.35$ and within a magnitude range $\Delta M\sim 7$ $mag$. The sample selection procedure is based on the Bernardi et al. (2003a) and Hyde \& Bernardi (2009) selection criteria. These criteria are as follows:

\begin{itemize}

\item The brightness profile of the galaxy must be well adjusted by a de Vaucouleurs profile, in both the $g$ and $r$ filters (fracdevg = 1 and fracdevr = 1 according to the SDSS nomenclature).

\item Galaxies must have an early-type spectrum ($eclass < 0$, according to the SDSS nomenclature).

\item We select objects with a de Vaucouleurs magnitude $14.5 < m_{r, dev} < 17.5$ and its equivalent in the g filter.

\item The quotient of the semi axes (b/a) for the ETGs must be larger than 0.6 in both filters $g$ and $r$.

\item We select objects with a velocity dispersion of $\sigma_0 >$ 60 $km/s$ and a signal-to-noise ratio (S/N) $>$ 10.



\end{itemize}

In addition, we extract a volume-limited sample of approximately 17 000 ETGs with 0.04 $\leq\;z\;\leq$ 0.08 in the g and r-band filters. This subsample covers a magnitude range $<\Delta M>$ $\sim 4.5$ $mag$ ($-18.5 \ge M_{g} > -23.0$) in both filters and we refer to it as the homogeneous SDSS sample. In Fig. 1, we show the behaviour of the magnitude as a function of redshift for galaxies contained in the SDSS total sample. Vertical lines represent the limits of the $0.04 \leq\;z\;\leq$ 0.08 redshift interval where the homogeneous sample of the SDSS is contained. We note that within these limits there is a deficiency of galaxies for $M_{g} \gtrsim -20.0$, so we may affirm that $M_{g} = -20.0$ represents the approximate completeness limit of the homogeneous SDSS sample.

\subsection{Correction of the photometric and spectroscopic data}

Once we have compiled the sample of galaxies, it is necessary to perform a number of corrections to both the photometric and the spectroscopic data. We now list the performed corrections:

\begin{itemize}

\item Seeing correction: We use the seeing-corrected de Vaucouleurs parameters (total magnitude and effective radius) from the SDSS pipeline.

\item Poor sky-subtraction correction: It is well known that the SDSS photometric reduction underestimates the luminosity of bright galaxies, as described by Hyde \& Bernardi (2009) whose results we adopt here to correct our sample of ETGs.

\item Extinction correction: We use the extinction correction values from the SDSS pipeline.

\item K correction: We use the K correction values from Bernardi et al. (2003a) and apply them to our sample as follows:

\begin{equation}
k_{g}(z)\, \; =\, \; -5.261\; z^{1.197},
\end{equation}

\begin{equation}
k_{r}(z)\, \; =\, \; -1.271\; z^{1.023}.
\end{equation}

\item Cosmological dimming correction: We use the cosmological dimming correction of J\o rgensen et al. (1995a).

\item Evolution correction: Bernardi et al. (2003b) report that the more distant galaxies in their sample are brighter than those nearby.
We use their results and apply them to our sample of galaxies as follows:

\begin{equation}
ev_{g}(z)\, \; =\, \; +1.15\; z,
\end{equation}

\begin{equation}
ev_{r}(z)\, \; =\, \; +0.85\; z.
\end{equation}

\item Effective radius correction to the rest reference frame: Given that ETGs have colour gradients, their mean effective radii at longer wavelengths are smaller. To correct for this effect, we follow the procedure given in Hyde \& Bernardi (2009). The average corrections for the g and r filters are 0.32 kpc and 0.27 kpc, respectively.

\item Aperture correction to the velocity dispersion: The velocity dispersion for ETGs appears to have radial gradients. This causes that the velocity dispersion values given in the SDSS ($\sigma_{SDSS}$) depend on both the distance of the object and the size of the aperture used for the observations ($r_{ap}$). To correct our data to a system that is independent from both the distance and instrument used for the observations, we use an expression derived by J\o rgensen et al. (1995b) as follows:

\begin{equation}
\frac{\sigma_{cor}}{\sigma_{SDSS}}\, \; =\, \; \left(\frac{r_{ap}}{r_{e}/8}\right)^{0.04},
\end{equation}

where $r_{ap}$ = $1.5$ $arcsec$ and $r_{e}$ is the effective radius in arcsec.

\end{itemize}

The final errors in the FJR variables were obtained from the errors given in the SDSS, which were in turn propagated by considering the mathematical expressions of each of the corrections listed above.

\section{The Faber-Jackson relation}

\label{calFJRelation}

The estimate of the FJR coefficients may be strongly affected by both the fitting method and the choice of the independent variable used in the fit (\cite{iso90}). Biases may be even larger if there are measurement errors in the variables, if the errors are correlated, or if there is an intrinsic dispersion in the relation. The Bivariate Correlated Errors and Intrinsic Scatter Bisector ($BCES_{Bis}$) method (\cite{iso90}; \cite{akr96}) is a statistical model that takes into account the different sources of bias mentioned above. In this paper, we use the $BCES_{Bis}$ method to calculate the FJR coefficients.

\subsection{Slope evolution of the FJR as function of the magnitude range}

\label{evolutionFJR}

From the photometric parameters of the different galaxy samples, we calculate the FJR coefficients in different magnitude intervals, the errors in these coefficients, and the total intrinsic dispersion (see \cite{lab03}). According to La Barbera et al. (2003), it is necessary to know the error values of each one of the variables involved in the structural relations for the calculation of the total intrinsic dispersion. These errors come directly from the SDSS data (see Sect. 2), while the errors in the FJR coefficients were calculated by us following Akritas \& Bershady (1996) in $1\sigma$ intervals.

The FJR coefficients were calculated in both increasing-magnitude-intervals, as well as in narrow-magnitude-intervals. Among other things, this allows us to characterise the behaviour of the FJR coefficients with respect to the width of the magnitude range and the luminosity of galaxies within the magnitude range. In Tables 1-3, we present the results obtained for the SDSS sample in the $g$ and $r$ filters.

In Table 1, we can see that the intrinsic dispersion in the FJR ($\sigma_{FJR}$) for each fit changes appreciably each time we include brighter galaxies in the sample (upper magnitude cut-off). Apart from these changes, we can also see variations in the coefficients of the FJR. These variations are greater than the associated errors in most cases (the differences in the $B$ coefficient may be as great as 85\%). Figure 2 shows the variation in the $B$ coefficient as function of the upper magnitude cut-off. On the other hand, when we perform the FJR analysis by considering first the brightest galaxies in each sample and progressively include fainter galaxies (lower magnitude cut-off),
the behaviour of the FJR parameters is similar to that described above: both the intrinsic dispersion and the FJR coefficient-values change systematically as we increase the width of the magnitude interval (see Table 2 and Fig. 3).

We also analyse the data by considering galaxy samples in progressively brighter fixed-width magnitude intervals  (narrow-magnitude-intervals), that is, by considering magnitude intervals of the same width over the whole magnitude range. The behaviour of the $B$ coefficient may be seen in Table 3 and Fig. 4, where we show that the differences in the $B$ coefficients in 1-$mag$ wide interval may be as great as 58\%. However, when the width of the magnitude interval is increasingly reduced, the differences from one magnitude range to another tend to disappear. That is, at constant magnitude ($\Delta M = 0$) the $B$ coefficient seems to be the same for both bright and faint galaxies.

To investigate whether the changes in the $B$ coefficients for increasing-magnitude-intervals and 1-$mag$ wide intervals are significant or whether they are only the product of statistical fluctuations, it is necessary to apply non-parametric tests to the data for the galaxy sample. The non-parametric tests appropriate to our study include the run test (\cite{ben66}; \cite{nig08}; \cite{nig09}). In Table 4, we show the results of the application of the run test method to the data. As a null hypothesis, we assume that there is no underlying trend in the data. The percentages given in Table 4 refer to the confidence level with which we may reject the null hypothesis. In Table 4, we may see that, on average, the null hypothesis may be rejected with a confidence level of 93 per cent. This implies that there are strong reasons to affirm that there is an underlying trend in the values of the $B$ coefficient of the FJR.


\subsection{Dependence of the structural relations on the magnitude range}

\label{comparisoFJRkrfp}

Nigoche-Netro (2007) and Nigoche-Netro et al. (2007; 2008; 2009) demonstrated, using both observational data and Monte Carlo simulations that the coefficients and the intrinsic dispersion of the KR and the FP depend on the width of the magnitude range and the luminosity of galaxies within this magnitude range. This dependence is caused by a `geometrical effect' because the distribution of the ETGs on the $log(R_e)$, $\left\langle \mu\right\rangle _{e}$ and $log(\sigma_0)$ parameter space depends on luminosity (magnitude segregation, see also \cite{des07}) and the geometrical form of the ETGs distribution in this space is not symmetrical (variations in the intrinsic dispersion). Since the FJR is a projection of the FP, the dependence of the values of the FJR coefficients on the luminosity and width of the magnitude range might be caused by a similar `geometrical effect' of the ETGs distribution in the plane of the variables $M$ and $log(\sigma_0)$. In fact, this is the case given that the distribution of the galaxies that define the FJR depends explicitly on luminosity. In Sect. 3.1, we established that the coefficients values and the intrinsic dispersion in the FJR change systematically as we consider brighter galaxies. This implies that the geometrical shape of the galaxy distribution changes as brighter galaxies are considered. This may easily be checked in Fig. 5 where one can see that as brighter galaxies are considered the shape of the galaxy distribution changes systematically.

From the aforementioned results, we may infer that the FJR coefficients and its intrinsic dispersion depend on the width of the magnitude range and the luminosity of galaxies within this magnitude range. This dependence is caused by a `geometrical effect', in other words,  the geometrical shape of the distribution of ETGs on the $M-log(\sigma_0)$ plane changes systematically as we consider brighter ETGs. The values of the FJR parameters (zero point, slope and intrinsic dispersion) are also affected, because the fitting of a straight line to a set of data does not provide the same result for data distributed with a rectangular shape as for data distributed with a triangular shape or, for that matter, with any other geometrical shape. In this sense, any other systematic restrictions imposed on a sample of ETGs, such as velocity dispersion cuts, will cause changes in the geometrical form of the distribution of ETGs on the $M-log(\sigma_0)$ plane. The more stringent these changes are, the more pronounced will be the changes in the values of the FJR coefficients.

It is important to point out here that when the magnitude interval is sufficiently narrow ($\Delta M < 1.0$ $mag$), the slope $B$ of the FJR is similar for both faint and bright galaxies. This behaviour is also found for the slopes of the KR and FP, which means that, at constant magnitude ($\Delta M = 0$) the slopes of the structural relations should be the same for faint and bright galaxies. However, there would be a difference in the values of the zero point for faint and bright galaxies. This difference would be proportional to the difference in luminosity (see also, \cite{jor96}; \cite{ben92}; \cite{ber03c}).

\section{Discussion}

Given the results presented in this article and those in Nigoche-Netro et al. (2008; 2009), it is important to recapitulate and summarise the implications that the dependence of the structural relations on the magnitude range has on the properties of ETGs.

Several papers have mentioned that the intrinsic dispersion of the FP is small ($\sim$ 0.1 dex) (\cite{kja93}; \cite{jor96}; \cite{kel97}; \cite{jor99}; \cite{bla02}; \cite{ber03c}; \cite{red05}; \cite{jor06}), although other studies demonstrate that the intrinsic dispersion is far from being small ($\sim0.3$ dex) (\cite{ben92}; \cite{lab03}; \cite{nig09}). This dispersion causes the galaxy distribution in the space defining the FP to follow a surface whose width is determined by this dispersion. For the KR and the FJR, something similar occurs, that is to say, the galaxy distribution follows a band whose width is determined by the intrinsic dispersion.

At this point, it is clear that the intrinsic dispersion, magnitude segregation, and the observational biases (or those biases caused by arbitrary cuts applied to the galaxy samples) are mainly responsible for the geometrical form of the galaxy distribution in the $log(R_e)$, $\left\langle \mu\right\rangle _{e}$ and $log(\sigma_0)$ space and that this geometrical form is the one that determines the coefficients of the structural relations.

In previous sections and Nigoche-Netro et al. (2008; 2009) we demonstrated that magnitude cuts (or cuts in any other of the variables involved in the structural relations) cause changes in the coefficients values of the structural relations and the more pronounced the cuts, the more pronounced the changes will be. It has also been shown that when the width of the magnitude interval diminishes, the differences in the slope of the structural relations (for intervals of the same width and different luminosity) become small and when we have approximately constant magnitude ($\Delta M \sim 0$) the differences are negligible (see Fig. 4). This behaviour has been labelled by the expression `geometrical effect'.

It is risky to draw
conclusions about the physical properties of galaxies by
comparing the slopes of the structural relations for magnitude
ranges of different widths or for magnitude ranges of the same
width but of different luminosity, because, with the exception of
the full magnitude interval, there is no ideal width at which
comparisons should be made. This means that, if the magnitude range is sufficiently narrow ($\Delta M < 1.0$ $mag$, see Fig. 4) the differences are negligible, whereas if it is wider ($\Delta M \geq 1.0$ $mag$, see Fig. 2) the geometrical form is dominated by the magnitude cut, which could mask any differences caused by the intrinsic physical properties of the galaxies. For samples of galaxies in magnitude ranges of the same width and luminosity but for different wavelengths, redshifts, or environments, comparison of the slope values could be useful in investigating whether these samples depend on the afore mentioned variables. However, the values of the slopes will only be representative
for the width of the interval under consideration. In other words, the value of the slope for a sample of galaxies within a specific magnitude width range may not be representative of the
entire magnitude range, nor of a different galaxy sample with a magnitude range of a different width.

In Sect. 1, we present a summary of the most significant works regarding the universality of the structural relations published in recent years. There are important discrepancies in terms of results. We present a list of the differences we have encountered:

\begin{itemize}

\item For the wavelength case, several studies indicate that the coefficients of the structural relations remain stable when considering different wavelengths (\cite{ben92}; \cite{BENDERETAL98}; \cite{ber03b}; \cite{ber03c}; \cite{lab05}; \cite{lab08}). However, other studies show that the structural relations depend on wavelength (\cite{jor96}; \cite{hud97}; \cite{pah98}; \cite{sco98}; \cite{jun08}).

\item As far as the environment is concerned, several studies imply that the structural relations and/or the structural parameters of galaxies are affected by the environment (\cite{ben92}; \cite{BENDERETAL98}; \cite{tru01}; \cite{tru02}; \cite{ber03b}; \cite{agu04}; \cite{gut04}; \cite{den05}; \cite{jor05}), although other studies reach exactly the opposite conclusion (\cite{ros01}; \cite{tre01}; \cite{evs02}; \cite{gon03}; \cite{red04}; \cite{red05}; \cite{nig07}).

\item When structural relations of galaxy samples are studied at different redshifts, several studies indicate that only the zero point of these relations depends on the redshift (\cite{bar98}; \cite{zie99}; \cite{lab03}; \cite{bar06}). Other authors find that there is a dependence of the zero point on redshift but that the slopes of the structural relations are steeper for higher redshift galaxies than for galaxies in the local Universe (\cite{tre05}; \cite{jor06}; \cite{fri09}).

\item If we consider luminosity, some authors find that dwarf and bright ellipticals follow structural relations with different coefficients (\cite{kor85}; \cite{ham87}; \cite{ben92}; \cite{cao93}; \cite{agu09}; \cite{des07}). Studies by Nigoche-Netro (2007), Nigoche-Netro et al. (2008), and Nigoche-Netro et al. (2009) as well as this paper find that the detected changes in the values of the coefficients of the structural relations obtained from fits of samples of galaxies of different luminosity are not necessarily caused by differences in the intrinsic properties of these galaxies.



\end{itemize}

Given that in the vast majority of the investigations carried out to date the effect produced by the magnitude restrictions on the ETGs structural relations has not been taken into consideration, we are able to establish that, at least, part of the discrepancies mentioned above may be precisely due to this effect.




We note that if there were differences in the total
intrinsic dispersion for faint and bright galaxies and we were
comparing magnitude ranges of gradually decreasing width. This process consists in separating the
sample in magnitude ranges of the same width and comparing them at
different luminosity, then, separating the sample in magnitude
ranges of smaller width and comparing them again at different
luminosity and so on, until a minimum width interval is reached. The differences obtained in this manner
would not disappear and when $\Delta M = 0$, we would precisely determine the intrinsic dispersions
for each of the magnitudes under study. This would occur because as a magnitude
interval becomes smaller, the geometrical form of the galaxy
distribution approximates that of a straight line. This quasi-linear
distribution would have a similar slope value at any luminosity, but
its `size' would depend on the luminosity being
considered (given that this is the initial premise). We therefore, consider
the most appropriate procedure for obtaining physical
information about a sample of galaxies would be to find its intrinsic
dispersion for each magnitude value and then perform comparisons of
this dispersion at different luminosity, wavelengths, redshifts, or environments.




Finally, it is important to mention that the `homogeneous' sample used in this article and those in Nigoche-Netro et al. (2008; 2009) are relatively complete in the bright part ($M_{g} \lesssim -20.0$) so in each one of them we are able to compare the intrinsic dispersion in different magnitude values. To a first approximation, Table 3 of this paper reveals that in the regime $M_{g}\lesssim -20.0$ the intrinsic dispersion for bright galaxies is smaller than that for faint galaxies (see also \cite{jor96}; \cite{hyd09}). Up to this moment, there is no conclusive explanation in the literature of this effect. It is imperative to investigate which are the physical processes responsible for the variation of the intrinsic dispersion values of the structural relations for bright and faint galaxies, as well as the exact limits of the galaxy distribution and the physical properties responsible for these limits. These topics will be discussed in a forthcoming paper.

\section{Conclusions}
\label{sec:conclusions}

Analysing the FJR as a function of the magnitude range we obtain the following:

\begin{itemize}

\item The parameters of the FJR depend on the magnitude range within which the galaxies of the sample under analysis are distributed.

\item The dependence of the FJR on the magnitude range may be explained by a `geometrical effect' (see Sects. 3.2 and 4 for full details).

\end{itemize}

With the results given above and the data from Nigoche-Netro et al. (2008; 2009), we find that the intrinsic dispersion, magnitude segregation and, to a larger extent, the observational biases (or those biases caused by arbitrary cuts made on galaxy samples) are
mainly responsible for the geometrical form of the galaxy distribution and that this form determines the values of the coefficients of the structural relations.

We find that it is risky to draw conclusions about the physical properties of galaxies by comparing the slopes of the structural relations in magnitude ranges of different widths or in magnitude ranges of the same width but of different luminosity, because, with the exception of the full magnitude interval, there is no privileged width for making comparisons. This means that, if the magnitude range is narrow the differences are negligible, whereas if it is wide, the geometrical form is dominated by the magnitude cut that could mask any differences caused by intrinsic physical properties of the galaxies. In the case of samples of galaxies in magnitude ranges of the same width and luminosity but of different wavelengths, redshifts, or environments, comparison of the slope values is also a delicate matter, since the magnitude cut could mask the intrinsic physical properties of the galaxies and the conclusions about these physical properties might be misinterpreted.

We find that, if there were differences in the total intrinsic dispersion for faint and bright galaxies and we were to compare magnitude ranges decreasing in width, the differences would not disappear (as happens with the differences for the slopes of the structural relations) and when $\Delta M = 0$, we would find the exact values for the intrinsic dispersions for each one of the magnitudes under study. We, therefore, consider the most appropriate procedure for obtaining physical information for a sample of galaxies to be finding its intrinsic dispersion at each magnitude value and then performing comparisons of this dispersion at different luminosities, wavelengths, redshifts, or environments. A first approach to the study of galaxies of different luminosities (see Sect. 4) reveals that in the regime $M_{g}\lesssim -20.0$ the intrinsic dispersion for bright galaxies is smaller than that for faint galaxies, confirming previous results by J\o rgensen et al. (1996) and Hyde \& Bernardi (2009). The study of the intrinsic dispersion as a function of luminosity requires a far more complete and detailed analysis. This analysis shall be published in a forthcoming paper.


We may conclude that part of the discrepancies found in the literature about the behaviour of the structural relations with respect to the following variables: wavelength, environment, redshift, and luminosity (see Sects. 1 and 4) must be due to the vast majority of the investigations carried out to date having underestimated the effects produced by magnitude restrictions as well as those restrictions that affect each one of the variables involved in the parameters of the structural relations. Finally, it is very important to redirect efforts towards investigating the physical processes that cause the intrinsic dispersion in the structural relations and also to bright galaxies having a different intrinsic dispersion from faint galaxies.




\section*{Acknowledgments}

We would like to dedicate this humble work to the memory of Mrs. Eutiquia Netro Castillo, an extraordinary woman.

We would like to thank Consejo Nacional de Ciencia y Tecnolog\'{\i}a (M\'{e}xico) for a postdoctoral fellowship number 81937, Instituto de Astrof\'isica de Canarias (IAC, Espa\~na) and Instituto de Astronom\'ia (UNAM, M\'exico) for all the facilities provided
for the realisation of this project. Partially
funded by the Spanish MEC under the Plan Nacional
de I+D 2007 grant AYA2007-67965-C03-01: Estallidos
(http://www.iac.es/project/GEFE/estallidos). P.L. acknowledges
the postdoctoral grant of the Spanish MEC within Estallidos. We thank Jana Benda for her expedient help with this paper. We express our deepest appreciation to the anonymous referee whose comments and suggestions greatly improved
the presentation of this paper.

\begin{table*}

\begin{minipage}[t]{\columnwidth}

\renewcommand{\footnoterule}{}  

\caption{Coefficients of the FJR for the SDSS sample of galaxies in increasing-magnitude-intervals (upper magnitude cut-off).}

\vskip1.0cm

\begin{tabular}{cccccc}

\hline
$MI$   \footnote{Absolute magnitude interval within which the galaxies are distributed.}         & $N$  \footnote{Number of galaxies in the magnitude interval.} & $B$ \footnote{Slope of the FJR.}              & $A$   \footnote{Zero point of the FJR.}     & $\sigma_{FJR}$ \footnote{Total intrinsic dispersion of the FJR.}\\

\hline


\multicolumn{5}{c}{}\\
\multicolumn{5}{c}{Total SDSS sample (g filter)}\\
\multicolumn{5}{c}{}\\

$-17.5 \geq M > -18.5$ & 111   & 0.695 $\pm$ 0.049 & -10.724 $\pm$ 0.898 & 0.235 \\
$-17.5 \geq M > -19.5$ & 1359  & 0.509 $\pm$ 0.027 & -7.669  $\pm$ 0.511 & 0.348 \\
$-17.5 \geq M > -20.5$ & 11421 & 0.345 $\pm$ 0.017 & -4.762  $\pm$ 0.338 & 0.397 \\
$-17.5 \geq M > -21.5$ & 40876 & 0.227 $\pm$ 0.012 & -2.489  $\pm$ 0.251 & 0.486 \\
$-17.5 \geq M > -22.5$ & 78237 & 0.166 $\pm$ 0.009 & -1.255  $\pm$ 0.203 & 0.571 \\
$-17.5 \geq M > -23.5$ & 88983 & 0.148 $\pm$ 0.009 & -0.891  $\pm$ 0.197 & 0.613 \\
$-17.5 \geq M > -24.5$ & 89308 & 0.147 $\pm$ 0.009 & -0.867  $\pm$ 0.198 & 0.617 \\


\multicolumn{5}{c}{}\\
\multicolumn{5}{c}{Total SDSS sample (r filter)}\\
\multicolumn{5}{c}{}\\
$-18.0 \geq M > -19.0$  & 60   &  0.930 $\pm$ 0.037  & -15.525 $\pm$ 0.685 & 0.225 \\
$-18.0 \geq M > -20.0$  & 878 &  0.509 $\pm$ 0.033  & -7.948 $\pm$ 0.647 & 0.328 \\
$-18.0 \geq M > -21.0$  & 7961 &  0.359 $\pm$ 0.019  & -5.269 $\pm$ 0.394 & 0.391 \\
$-18.0 \geq M > -22.0$  & 32969 &  0.239 $\pm$ 0.013  & -2.889 $\pm$ 0.028 & 0.465 \\
$-18.0 \geq M > -23.0$  & 71327 &  0.169 $\pm$ 0.009  & -1.446 $\pm$ 0.213 & 0.544 \\
$-18.0 \geq M > -24.0$  & 88224 &  0.145 $\pm$ 0.008  & -0.927 $\pm$ 0.195 & 0.589 \\
$-18.0 \geq M > -25.0$  & 89305 &  0.142 $\pm$ 0.008  & -0.868 $\pm$ 0.197 & 0.598 \\

\multicolumn{5}{c}{}\\
\multicolumn{5}{c}{ Homogeneous SDSS sample (g filter)}\\
\multicolumn{5}{c}{}\\
$-18.5 \geq M > -19.5$  & 739   &  0.783 $\pm$ 0.019  & -12.986 $\pm$ 0.384 & 0.253 \\
$-18.5 \geq M > -20.5$  & 7515 &  0.427 $\pm$ 0.015  & -6.404 $\pm$ 0.304 & 0.343 \\
$-18.5 \geq M > -21.5$  & 15078 &  0.235 $\pm$ 0.016  & -2.609 $\pm$ 0.336 & 0.479 \\

$-18.5 \geq M > -22.5$  & 16887 &  0.197 $\pm$ 0.017  & -1.859 $\pm$ 0.347 & 0.529 \\
$-18.5 \geq M > -23.0$  & 16901 &  0.196 $\pm$ 0.017  & -1.849 $\pm$ 0.349 & 0.531 \\

\multicolumn{5}{c}{}\\
\multicolumn{5}{c}{ Homogeneous SDSS sample (r filter)}\\
\multicolumn{5}{c}{}\\
$-19.0 \geq M > -20.0$  & 396   &  0.814 $\pm$ 0.027  & -14.022 $\pm$ 0.539 & 0.245 \\
$-19.0 \geq M > -21.0$  & 5617 &  0.472 $\pm$ 0.015  & -7.599 $\pm$ 0.315 & 0.320 \\
$-19.0 \geq M > -22.0$  & 13905 &  0.245 $\pm$ 0.017  & -2.983 $\pm$ 0.349 & 0.452 \\
$-19.0 \geq M > -23.0$  & 16828 &  0.192 $\pm$ 0.016  & -1.879 $\pm$ 0.349 & 0.515 \\
$-19.0 \geq M > -23.5$  & 16910 &  0.189 $\pm$ 0.016  & -1.829 $\pm$ 0.354 & 0.519 \\

\hline

\end{tabular}
\end{minipage}
\end{table*}

\begin{table*}

\begin{minipage}[t]{\columnwidth}

\renewcommand{\footnoterule}{}  

\caption{Coefficients of the FJR for the SDSS sample of galaxies in increasing-magnitude-intervals (lower magnitude cut-off).}

\vskip1.0cm

\begin{tabular}{cccccc}

\hline

$MI$   \footnote{Absolute magnitude interval within which the galaxies are distributed.}         & $N$  \footnote{Number of galaxies in the magnitude interval.} & $B$ \footnote{Slope of the FJR.}              & $A$   \footnote{Zero point of the FJR.}     & $\sigma_{FJR}$ \footnote{Total intrinsic dispersion of the FJR.}\\

\hline


\multicolumn{5}{c}{}\\
\multicolumn{5}{c}{Total SDSS sample (g filter)}\\
\multicolumn{5}{c}{}\\

$-23.5 \geq M > -24.5$ & 325   & 1.082 $\pm$ 0.314 & -23.132 $\pm$ 7.439 & 0.154 \\
$-22.5 \geq M > -24.5$ & 11071  & 0.441 $\pm$ 0.014 & -7.647  $\pm$ 0.311 & 0.267 \\
$-21.5 \geq M > -24.5$ & 48432 & 0.242 $\pm$ 0.011 & -2.999  $\pm$ 0.246 & 0.406 \\
$-20.5 \geq M > -24.5$ & 77887 & 0.173 $\pm$ 0.009 & -1.449  $\pm$ 0.208 & 0.532 \\
$-19.5 \geq M > -24.5$ & 87949 & 0.152 $\pm$ 0.009 & -0.976  $\pm$ 0.193 & 0.599 \\
$-18.5 \geq M > -24.5$ & 89197 & 0.148 $\pm$ 0.009 & -0.884  $\pm$ 0.195 & 0.615 \\
$-17.5 \geq M > -24.5$ & 89308 & 0.147 $\pm$ 0.009 & -0.867  $\pm$ 0.198 & 0.617 \\


\multicolumn{5}{c}{}\\
\multicolumn{5}{c}{Total SDSS sample (r filter)}\\
\multicolumn{5}{c}{}\\
$-24.0 \geq M > -25.0$  & 1081   &  0.525 $\pm$ 0.033  & -10.232 $\pm$ 0.796 & 0.168 \\
$-23.0 \geq M > -25.0$  & 17978 &  0.297 $\pm$ 0.015  & -4.548 $\pm$ 0.2362 & 0.295 \\
$-22.0 \geq M > -22.0$  & 56336 &  0.196 $\pm$ 0.011  & -2.113 $\pm$ 0.248 & 0.431 \\
$-21.0 \geq M > -25.0$  & 81344 &  0.157 $\pm$ 0.009  & -1.208 $\pm$ 0.205 & 0.538 \\
$-20.0 \geq M > -25.0$  & 88427 &  0.145 $\pm$ 0.008  & -0.938 $\pm$ 0.192 & 0.585 \\
$-19.0 \geq M > -25.0$  & 89245 &  0.143 $\pm$ 0.008  & -0.876 $\pm$ 0.195 & 0.596 \\
$-18.0 \geq M > -25.0$  & 89305 &  0.142 $\pm$ 0.008  & -0.868 $\pm$ 0.197 & 0.598 \\

\multicolumn{5}{c}{}\\
\multicolumn{5}{c}{Homogeneous SDSS sample (g filter)}\\
\multicolumn{5}{c}{}\\

$-22.5 \geq M > -23.0$ & 14   & 1.038 $\pm$ 0.206 & -20.988 $\pm$ 4.675 & 0.056 \\
$-21.5 \geq M > -23.0$ & 1823  & 0.505 $\pm$ 0.022 & -8.666  $\pm$ 0.487 & 0.218 \\
$-20.5 \geq M > -23.0$ & 9386 & 0.287 $\pm$ 0.019 & -3.796  $\pm$ 0.397 & 0.371 \\
$-19.5 \geq M > -23.0$ & 16162 & 0.211 $\pm$ 0.016 & -2.138  $\pm$ 0.339 & 0.498 \\
$-18.5 \geq M > -23.0$ & 16901 & 0.196 $\pm$ 0.017 & -1.849  $\pm$ 0.349 & 0.531 \\


\multicolumn{5}{c}{}\\
\multicolumn{5}{c}{Homogeneous SDSS sample (r filter)}\\
\multicolumn{5}{c}{}\\
$-23.0 \geq M > -23.5$  & 82   &  0.629 $\pm$ 0.071  & -12.142 $\pm$ 1.647 & 0.109 \\
$-22.0 \geq M > -23.5$  & 3005 &  0.376 $\pm$ 0.026  & -6.084 $\pm$ 0.571 & 0.255 \\
$-21.0 \geq M > -23.5$  & 11293 &  0.242 $\pm$ 0.019  & -2.989 $\pm$ 0.404 & 0.398 \\
$-20.0 \geq M > -23.5$  & 16514 &  0.197 $\pm$ 0.016  & -2.003 $\pm$ 0.346 & 0.498 \\
$-19.0 \geq M > -23.5$  & 16910 &  0.189 $\pm$ 0.016  & -1.829 $\pm$ 0.354 & 0.519 \\

\hline

\end{tabular}
\end{minipage}
\end{table*}

\begin{table*}

\begin{minipage}[t]{\columnwidth}

\renewcommand{\footnoterule}{}  

\caption{Coefficients of the FJR for the SDSS sample in narrow-magnitude-intervals (1-$mag$ wide interval).}

\vskip1.0cm

\begin{tabular}{cccccc}

\hline
$MI$   \footnote{Absolute magnitude interval within which the galaxies are distributed.}         & $N$  \footnote{Number of galaxies in the magnitude interval.} & $B$ \footnote{Slope of the FJR.}              & $A$   \footnote{Zero point of the FJR.}     & $\sigma_{FJR}$ \footnote{Total intrinsic dispersion of the FJR.}\\
\hline


\multicolumn{5}{c}{}\\
\multicolumn{5}{c}{Total SDSS sample (g filter)}\\
\multicolumn{5}{c}{}\\

$-17.5 \geq M > -18.5$ & 111   & 0.695 $\pm$ 0.049 & -10.724 $\pm$ 0.898 & 0.235 \\
$-18.5 \geq M > -19.5$ & 1248  & 0.725 $\pm$ 0.016 & -11.843  $\pm$ 0.301 & 0.256 \\
$-19.5 \geq M > -20.5$ & 10062 & 0.597 $\pm$ 0.008 & -9.864  $\pm$ 0.166 & 0.252 \\
$-20.5 \geq M > -21.5$ & 29455 & 0.506 $\pm$ 0.006 & -8.407  $\pm$ 0.123 & 0.267 \\
$-21.5 \geq M > -22.5$ & 37361 & 0.451 $\pm$ 0.006 & -7.576  $\pm$ 0.134 & 0.266 \\
$-22.5 \geq M > -23.5$ & 10746 & 0.526 $\pm$ 0.009 & -9.582  $\pm$ 0.226 & 0.230 \\
$-23.5 \geq M > -24.5$ & 325 & 1.082 $\pm$ 0.314 & -23.132  $\pm$ 7.439 & 0.154 \\


\multicolumn{5}{c}{}\\
\multicolumn{5}{c}{Total SDSS sample (r filter)}\\
\multicolumn{5}{c}{}\\
$-18.0 \geq M > -19.0$  & 60   &  0.930 $\pm$ 0.037  & -15.525 $\pm$ 0.685 & 0.225 \\
$-19.0 \geq M > -20.0$  & 818 &  0.715 $\pm$ 0.021  & -12.019 $\pm$ 0.403 & 0.253 \\
$-20.0 \geq M > -21.0$  & 7083 &  0.608 $\pm$ 0.009  & -10.434 $\pm$ 0.193 & 0.248 \\
$-21.0 \geq M > -22.0$  & 25008 &  0.509 $\pm$ 0.006  & -8.745 $\pm$ 0.133 & 0.268 \\
$-22.0 \geq M > -23.0$  & 38358 &  0.424 $\pm$ 0.006  & -7.219 $\pm$ 0.143 & 0.266 \\
$-23.0 \geq M > -24.0$  & 16897 &  0.396 $\pm$ 0.011  & -6.846 $\pm$ 0.254 & 0.242 \\
$-24.0 \geq M > -25.0$  & 1081 &  0.525 $\pm$ 0.033  & -10.232 $\pm$ 0.796 & 0.168 \\

\multicolumn{5}{c}{}\\
\multicolumn{5}{c}{Homogeneous SDSS sample (g filter)}\\
\multicolumn{5}{c}{}\\

$-18.5 \geq M > -19.5$ & 739   & 0.783 $\pm$ 0.019 & -12.986 $\pm$ 0.384 & 0.253 \\
$-19.5 \geq M > -20.5$ & 6776  & 0.593 $\pm$ 0.009 & -9.772  $\pm$ 0.194 & 0.257 \\
$-20.5 \geq M > -21.5$ & 7563 & 0.485 $\pm$ 0.012 & -7.911  $\pm$ 0.247 & 0.262 \\
$-21.5 \geq M > -22.5$ & 1809 & 0.532 $\pm$ 0.020 & -9.253  $\pm$ 0.446 & 0.210 \\
$-22.5 \geq M > -23.0$ & 14 & 1.038 $\pm$ 0.206 & -20.988  $\pm$ 4.675 & 0.056 \\


\multicolumn{5}{c}{}\\
\multicolumn{5}{c}{Homogeneous SDSS sample (r filter)}\\
\multicolumn{5}{c}{}\\
$-19.0 \geq M > -20.0$  & 396   &  0.814 $\pm$ 0.027  & -14.022 $\pm$ 0.539 & 0.245\\
$-20.0 \geq M > -21.0$  & 5221 &  0.623 $\pm$ 0.010  & -10.727 $\pm$ 0.215 & 0.258 \\
$-21.0 \geq M > -22.0$  & 8288 &  0.471 $\pm$ 0.011  & -7.894 $\pm$ 0.246 & 0.266 \\
$-22.0 \geq M > -23.0$  & 2923 &  0.434 $\pm$ 0.021  & -7.362 $\pm$ 0.469 & 0.231 \\
$-23.0 \geq M > -23.5$  & 82 &  0.629 $\pm$ 0.071  & -12.142 $\pm$ 1.647 & 0.109 \\

\hline

\end{tabular}
\end{minipage}
\end{table*}

\clearpage
\begin{table*}

\begin{minipage}[t]{\columnwidth}

\renewcommand{\footnoterule}{}  

\caption{Run test for the evaluation of the $B$ coefficient of the FJR from the SDSS sample of galaxies.}

\vskip1.0cm

\begin{tabular}{lc}

\multicolumn{1}{c}{}\\
\hline

&\\

   Sample        & per cent \footnote{The null hypothesis establishes that there is no underlying trend in the data, the percentages refer to the confidence level with which we can reject the null hypothesis.}
  \\

&\\

 Increasing magnitude intervals (upper magnitude cut-off)   & \\

 \hline

&\\

Total SDSS ($g$ filter)& 96 \\
Total SDSS ($r$ filter)& 96 \\

&\\

1 $mag$ interval   & \\

 \hline
&\\

Total SDSS ($g$ filter)& 90 \\
Total SDSS ($r$ filter)& 90 \\

\hline

\end{tabular}
\end{minipage}
\end{table*}

\clearpage

\begin{figure*}

\centering
\includegraphics[angle=0,width=12cm]{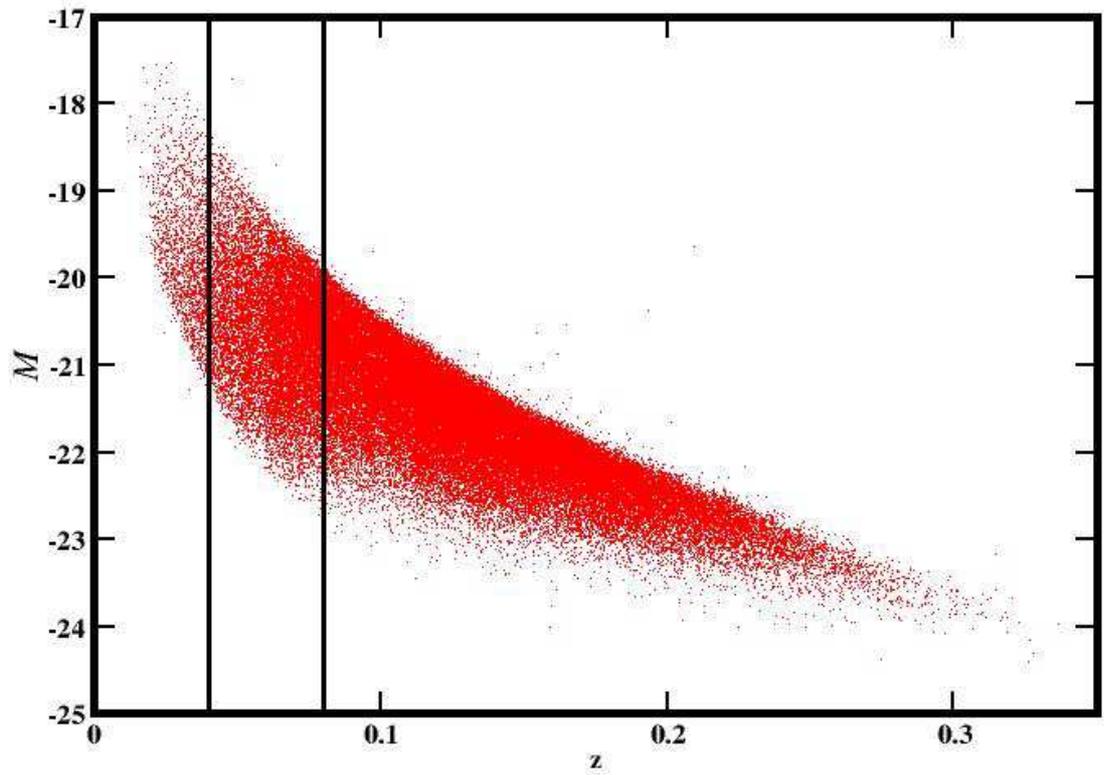}
\caption{Distribution of the absolute magnitude as function of redshift for the total SDSS sample (g filter). The vertical lines correspond to the 0.04 $\leq\;z\;\leq$ 0.08 redshift interval within which is contained the homogeneous subsample. In this interval, we note that for $M_{g} \gtrsim -20.0$ there exists a deficiency of galaxies, so $M_{g} = -20.0$ could be considered to be the completeness limit of the homogeneous SDSS sample.}

\end{figure*}

\clearpage

\begin{figure*}

\centering
\includegraphics[angle=0,width=12cm]{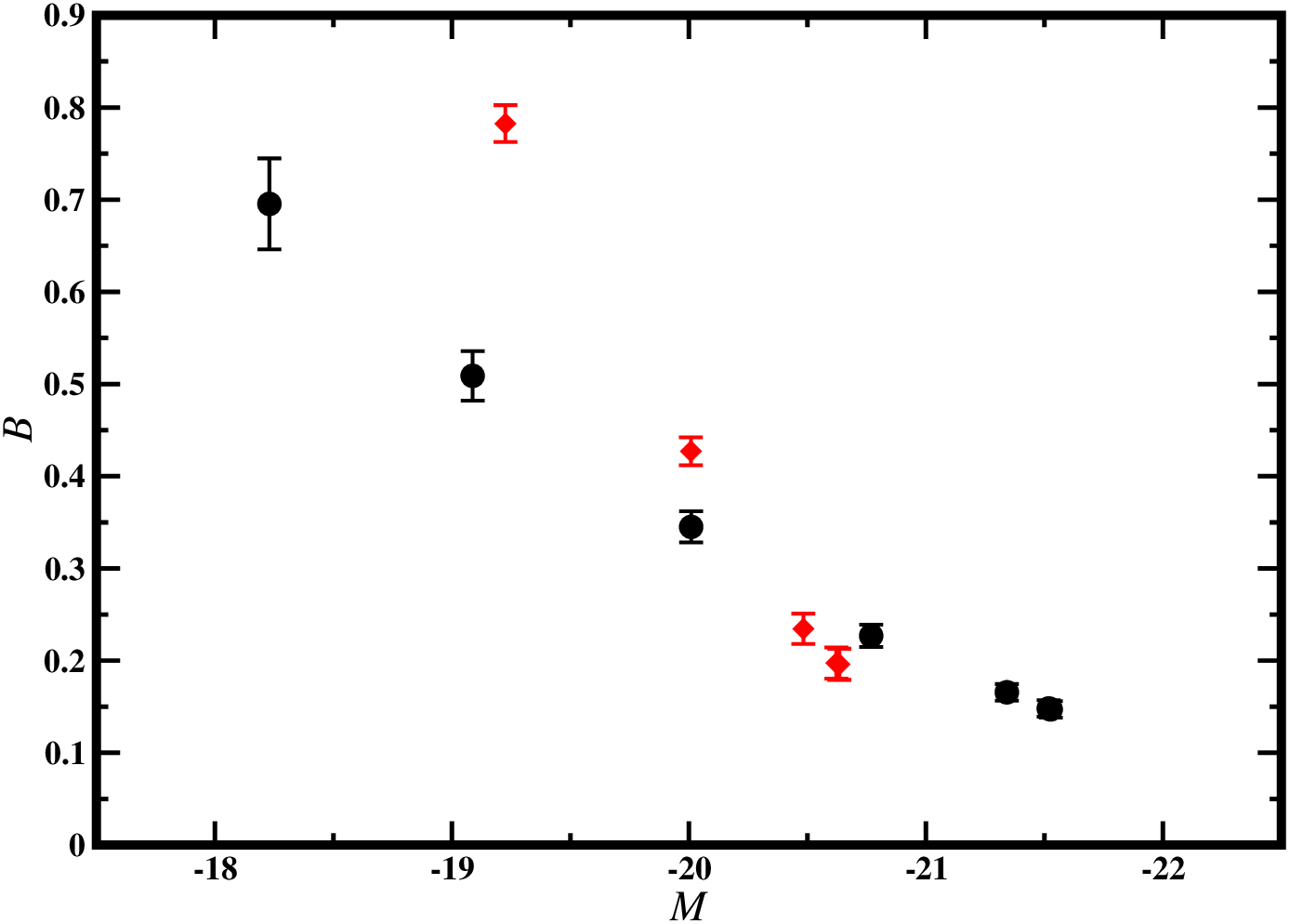}
\caption{Variation in the FJR slope ($B$) for increasing-magnitude-intervals (upper magnitude cut-off). Each point corresponds to the mean value of the total absolute magnitude of the galaxies contained in each magnitude interval analysed (see Table 1). Circles represent the total SDSS sample (g filter). Diamonds represent the homogeneous SDSS sample (g filter).}

\end{figure*}

\clearpage





\begin{figure*}

\centering
\includegraphics[angle=0,width=12cm]{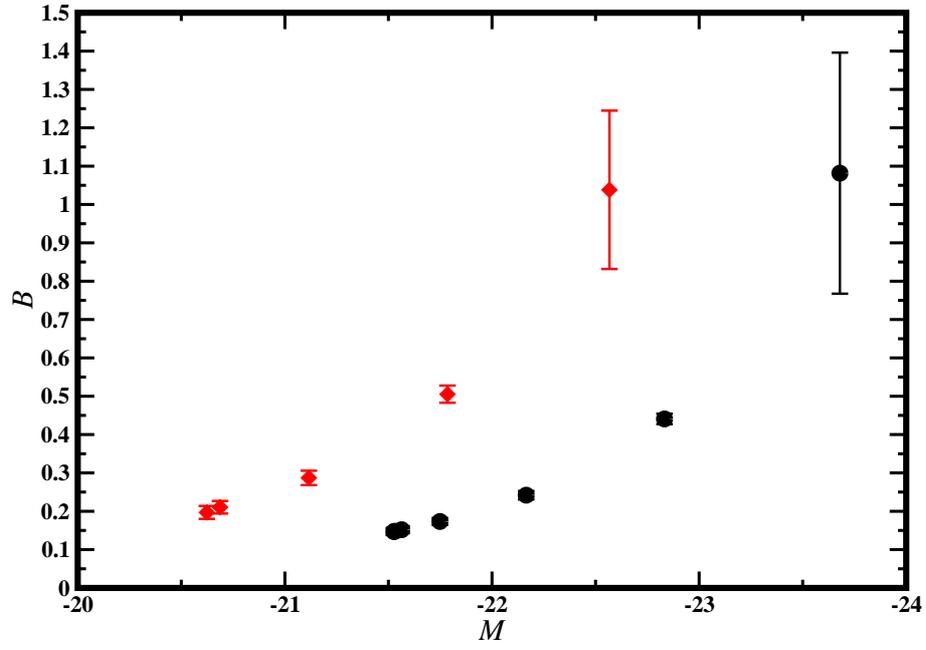}
\caption{Variation in the FJR slope ($B$) for increasing-magnitude-intervals (lower magnitude cut-off). Each point corresponds to the mean value of the total absolute magnitude of the galaxies contained in each magnitude interval analysed (see Table 2). Circles represent the total SDSS sample (g filter). Diamonds represent the homogeneous SDSS sample (g filter).}

\end{figure*}

\clearpage

\begin{figure*}

\centering
\includegraphics[angle=0,width=12cm]{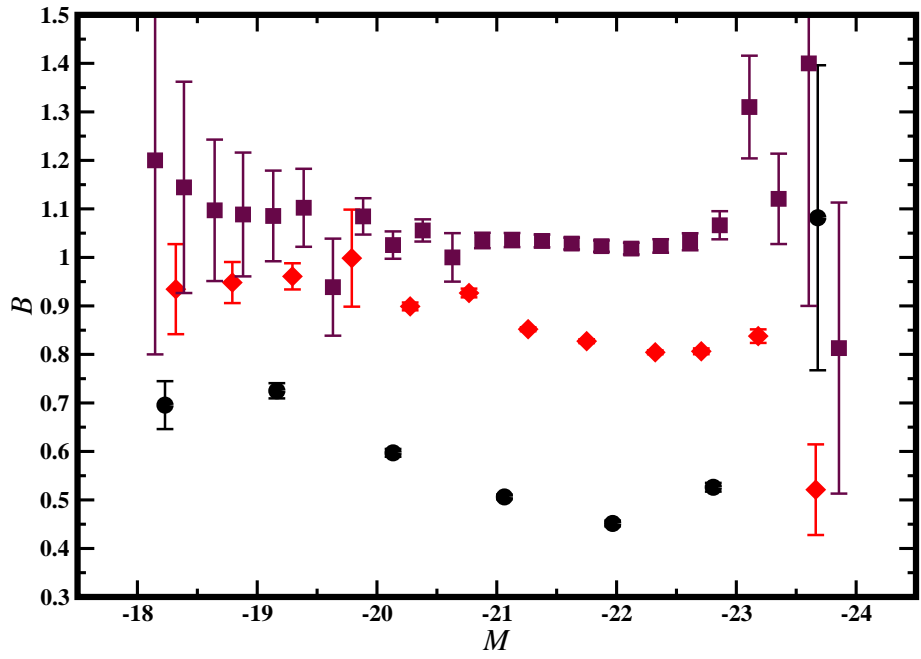}
\caption{Variation in the FJR slope ($B$) for narrow-magnitude-intervals. Each point corresponds to the mean value of the total absolute magnitude of the galaxies contained in each magnitude interval analysed. Circles represent the total SDSS sample ($g$ filter) in 1.0 $mag$ width intervals. Diamonds represent the total SDSS sample ($g$ filter) in 0.5 $mag$ width intervals. Squares represent the total SDSS sample ($g$ filter) in 0.25 $mag$ width intervals. }


\end{figure*}

\clearpage

\begin{figure*}

\centering
\includegraphics[angle=0,width=12cm]{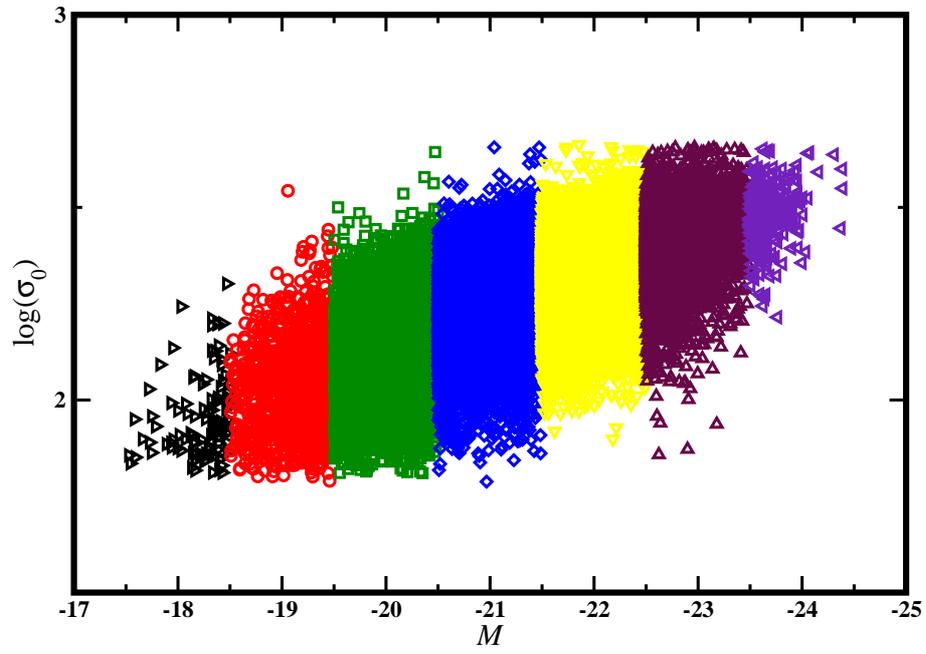}
\caption{Distribution of the galaxies in the total SDSS sample ($g$ filter) on the M-log($\sigma_0$) plane. Each symbol and colour represent a 1-$mag$ wide interval (see Table 3).}

\end{figure*}

\end{document}